Glass-Like Arrest in Spinodal Decomposition as a Route to Colloidal Gelation

S. Manley,[1,*] H. M. Wyss,[1] K. Miyazaki,[2,†] J. C. Conrad,[1] V. Trappe,[3] L. J. Kaufman,[2,†] D. R. Reichman,[2,†] and D. A. Weitz[1]

1. Dept. of Physics & DEAS, Harvard University, Cambridge, MA 02138
2. Dept. of Chemistry & Chemical Biology, Harvard University, Cambridge, MA 02138
USA
3. Dept. of Physics, University of Fribourg, CH-1700 Fribourg, Switzerland

* Current Address: Dept. of Chemical Eng., Massachusetts Institute of Technology, Cambridge, MA, 02139
† Current Address: Dept. of Chemistry, Columbia University, New York, NY, 10027

**Abstract**

Colloid-polymer mixtures can undergo spinodal decomposition into colloid-rich and colloid-poor regions. Gelation results when interconnected colloid-rich regions solidify. We show that this occurs when these regions undergo a glass transition, leading to dynamic arrest of the spinodal decomposition. The characteristic length scale of the gel decreases with increasing quench depth, and the nonergodicity parameter exhibits a pronounced dependence on scattering vector. Mode coupling theory gives a good description of the dynamics, provided we use the full static structure as input.

Colloid-polymer mixtures exhibit a rich phase behavior. Depletion of polymer between two colloidal particles induces an attractive interaction whose magnitude, $U$, is set by the polymer concentration in the free volume, $c_p$. The range of this interparticle attraction is determined by the radius of the polymer, $R_p$, and is expressed by $\xi = R_p/a$, with $a$ the radius of the colloidal particles. The phase behavior is controlled by both $c_p$ and $\xi$, as well as the colloid volume fraction, $\phi$. The particles can form equilibrium phases, including crystals, fluids and gases [1, 2]. Additionally, there are a wide variety of non-equilibrium states, characterized by the dynamic arrest of the colloidal particles [3-6]. In the absence of polymer, crowding causes the system to dynamically arrest into a repulsive glassy state at $\phi \sim 0.58$, where particles are caged by their nearest-neighbors. This dynamic arrest is captured by mode coupling theory (MCT), a mean-field-like theory which uses the static structure factor, $S(q)$, to predict the dynamics [7, 8]. Upon addition of a small amount of polymer, arrest is no longer driven by crowding, but is instead driven by weak attraction between particles; surprisingly, this attractive glass can also be described by MCT [4, 9]. In both cases, the dominant length scale in the structure corresponds to the nearest-neighbor particle separation. By contrast, at lower $\phi$ and higher $U$, the system can gel by forming a space-spanning network that is characterized by an additional, larger length scale. For systems at very low $\phi$ and sufficiently high $U$, this gelation is a consequence of kinetics; fractal clusters grow and ultimately form a gel by cluster-cluster aggregation [10, 11]. The theory for kinetic aggregation naturally includes the larger length scale, which is the cluster size. By contrast, at somewhat higher $\phi$ and lower $U$, phase separation such as spinodal decomposition can drive large-scale structuring, and thereby play an important role in determining the pathways for gel formation and the resultant network structure [12-14]. In this case, the mechanisms for arrest during gelation are still unclear. Gelation requires the formation of a solidified, space spanning network. In spinodal decomposition this is driven by a dynamic arrest within the connected colloid-rich region, which has been postulated to result from percolation [15-19], pinning [20] or a glass-like transition [6, 12, 13, 17, 21, 22]. An experimental investigation that distinguishes these possible mechanisms for dynamic arrest, and that determines the correct underlying physics is essential to provide guidance for a theoretical description of this class of gelation

In this Letter, we show that when gelation of colloid-polymer mixtures follows spinodal decomposition, solidification occurs by means of a glass transition within the interconnected colloid-rich phase. We calculate the dynamics with MCT [12], and include modes at both high and low wave vectors, $q$, using as input the structure factor, $S(q)$, determined by static light scattering (SLS). We find good agreement between the calculated $q$-dependent non-ergodicity factors, and the measured values determined by dynamic light scattering (DLS).

We use sterically stabilized poly-methylmethacrylate (PMMA) spheres, with radius $a = 136 \pm 5$ nm, as determined by DLS. Particles are suspended at a volume fraction of $\phi = 0.25$ in an index- and density-matching solvent mixture of cycloheptyl bromide and decalin. This choice of solvents allows us to probe the samples using both light scattering and microscopy, while minimizing the influence of gravity. Since the colloids are slightly charged in these solvents, samples are prepared with 1 mM tetra-butyl-ammonium chloride [23] to suppress any Coulombic contribution to the interaction potential. Attractive depletion interactions are induced by adding linear polystyrene with a molecular weight $M_w = 2 \times 10^6$ Daltons. Our samples are homogenized by slow mixing for at least 12 hours prior to measurement. To approximate the interaction potential between two particles separated by a distance $r$, we use the Asakura-Oosawa potential, $\frac{U}{k_B T} = \Pi(c_p) V(r)$, where $\Pi(c_p)$ is the osmotic pressure, and $V(r)$ is the overlap volume of their depletion zones [24, 25]. We determine $\Pi(c_p)$ by integrating the osmotic modulus, $d\Pi(c_p)/dc_p$, obtained with SLS from polystyrene solutions. In all cases, $c_p$ is below the overlap concentration. Because we use a good solvent for polystyrene, $R_p$ decreases with increasing $c_p$; accordingly $\xi$ decreases from ~ 0.3 to ~ 0.2.

We study the structures formed at different $U$ by measuring the static light scattering intensity, $I(q)$, over a wide range of scattering vectors. Using a goniometer with an Ar-ion laser, with a wavelength *in vacuo* $\lambda = 514.5$ nm, we measure $I(q)$ for 2.4 $\mu$m$^{-1}$ < $q$ < 21 $\mu$m$^{-1}$. We use the Percus-Yevick (PY) [26] approximation to calculate the static structure factor for hard spheres, $S_{HS}(q)$. We normalize the scattering intensity obtained from a sample without any polymer by $S_{HS}(q)$ to determine the particle form factor, $F(q)$, which we use to obtain $S(q)$ for all other samples. All data are affected by multiple scattering which increases the intensity near the minimum of the form factor [27], and the normalization used helps to mitigate this effect. To access lower $q$, 0.2 $\mu$m$^{-1}$ < $q$ < 1.7 $\mu$m$^{-1}$, the scattered light obtained using a collimated He-Ne laser, with $\lambda = 633$ nm, is projected onto a screen, which is imaged using a CCD camera. The intensity is averaged around rings of constant $q$, and normalized by a factor chosen to best match the low- and high-$q$ data for the sample with $c_p = 0$. This combination of measurements allows us to probe structures from length scales just below the single particle size, $qa \sim 4$, up to tens of particle diameters, $qa \sim 0.03$.

For weak attractions, $U \leq 2.6\ k_BT$, the dominant feature in $S(q)$ is the nearest-neighbor peak, as shown in Fig. 1. The peak position shifts to higher $q$ relative to that of $S_{HS}(q)$; this reflects that particles are, on average, closer together. Concomitantly, there is a slight increase in $S(q)$ at low $q$ by comparison to $S_{HS}(q)$. As we increase the attraction further, $U \geq 2.9\ k_BT$, the magnitude of the particle-particle peak increases sharply, and its width becomes narrower, reflecting a strong increase in correlations between neighboring particles. This increase is highlighted by $\Delta S_m$, the difference between the value of $S(q)$ at the nearest-neighbor peak and its value at the minimum at lower $q$, shown in the inset of Fig. 1. Concurrently, an intense peak appears at small wave vectors, reflecting the sudden appearance of large-scale structures. This peak appears almost immediately after mixing ceases, and shows only weak time-dependence during a 30-min. measurement, shifting to slightly lower $q$. The position of the low-$q$ peak shifts to higher $q$ as $U$ is increased, indicating that the characteristic length of the system becomes smaller; this is accompanied by a reduction in the degree of spatial correlation at the nearest-neighbor length scale, indicated by a decrease in $\Delta S_m$. Once the gels form, we see little evidence of aging in either their structure or dynamics.

To further elucidate the morphology of the network, we use coherent anti-Stokes Raman scattering (CARS) microscopy [28, 29] to directly image the samples. In this technique, two pulsed lasers are tuned such that the frequency difference between them is resonant with a Raman-active vibration present in the solvent molecules but absent in the particles. This creates optical contrast between the solvent and the particles without the need for a fluorophor, allowing us to directly image the samples used in the light scattering experiments. At low interaction energies, $U \leq 2.6\ k_BT$, there is no resolvable structure in the CARS images. However, upon a slight increase in the attraction, $U \geq 2.9\ k_BT$, large, space-spanning networks appear, filling the entire field of view, as shown in Fig. 2. The structures are static over the ~ 30 min. measurement time; thus, these images provide direct evidence of the presence of a dynamically arrested network that spans space. As $U$ increases, the characteristic length scale of the network decreases. This is qualitatively consistent with the $U$-dependence of the low-$q$ peak in $S(q)$; furthermore, spatial Fourier transforms of the images exhibit a peak at $q_c$ consistent with that measured with light scattering. The resultant characteristic sizes, $R_c \sim \pi/q_c$, are shown by the circles in Fig. 2. These structures are much larger than expected for diffusion-limited cluster aggregation [10], which predicts $R_c \sim 3a$, too small to be resolved in the CARS images.

The abrupt appearance of space-spanning networks with increasing $U$ is accompanied by a pronounced change in the dynamic structure factor, $f(q,t)$. For $U \leq 2.6\ k_BT$, $f(q,t)$ is ergodic, as shown by the lines in Fig. 3(a), where we plot $f(q,t)$ for $qa = 3.5$ as a function of the viscosity-corrected delay time, $t\eta_0/\eta$, with $\eta_0$ and $\eta$ the viscosities of the solvent and the background polymer solutions respectively. The decays are reasonably well described by single exponentials and the decay times increase only slightly with $U$. For $U \geq 2.9\ k_BT$, $f(q,t)$ no longer fully decays, as shown by the symbols in Fig. 3(a); this is consistent with dynamic arrest of the network over the time scales of the experiment. For these samples, we use the ensemble-averaged scattering intensity to correct the measured time-averaged data and to determine $f(q,t)$ [30].

The sudden onset of these dynamically arrested networks with increasing $U$ suggests that the system crosses the spinodal line, as shown schematically in the inset of

Fig. 2. Samples quenched above the spinodal line separate into interconnected colloid-poor and colloid-rich regions which span space. Gelation occurs when the local particle concentration in the colloid-rich regions is sufficiently high that the particles become dynamically arrested; this leads to a solidification of the network and arrest of the spinodal decomposition. In spinodal decomposition, the rate of increase of the local $\phi$ increases with quench depth, which depends on $U$; thus, dynamic arrest occurs more quickly for higher $U$, reducing the coarsening time of the characteristic length scale in the phase separating system. Likewise, the initial characteristic length scale in spinodal decomposition also decreases with quench depth. Thus, the characteristic length scale of the network decreases with increasing $U$, as shown in Figs. 1 and 2.

There are several mechanisms which may be responsible for the dynamic arrest within the colloid-rich region, including percolation [15-19], a cluster glass transition [6, 22], or an attractive glass transition [12, 13, 17, 21]. To help elucidate the mechanism for the arrest, we measure $f(q,t)$ as a function of $q$. Although the samples remain non-ergodic for all $q$, a slow decay persists even at the longest times measured, as shown for $U = 2.89$ $k_BT$ in Fig 3(b). Thus, we determine the non-ergodicity factors, $f_c(q)$, from the value of $f(q,t)$ at $t\eta_0/\eta = 0.2$ sec. For each $U$, $f_c(q)$ is not monotonic in $q$; there are pronounced increases both at low $q$ and at the nearest-neighbor peak, $qa \sim 3.5$, as shown by the symbols in Fig. 4; this is reminiscent of the shape of $S(q)$. These results cannot be explained by a percolation scenario, since dynamic arrest of the percolation type predicts that $f_c(q)$ decreases monotonically with $q$, and grows continuously from zero as $U$ increases [18, 19]. Instead, we compare the data with the behavior of $f(q,t)$ calculated from MCT. Although MCT cannot rigorously be applied to highly heterogeneous, non-equilibrium systems [31], it should nevertheless capture the qualitative nature of this gelation since the colloid-rich phase is solely responsible for the observed slow dynamics and dynamical arrest, and since there is a separation of time scales between local, glassy freezing and large scale restructuring of the gel [12]. However, to account for gelation, we must use the full $S(q)$ as input, including the peak at low $q$ which accounts for the larger length scale structures of the gel; thus, we use the measured $S(q)$ shown in Fig. 1. In addition, since the MCT results are sensitive to very high $q$ contributions of $S(q)$, we extend the data using values calculated with the Asakura-Oosawa potential [32]. The volume fraction of the colloid-rich region, $\tilde{\phi}$, is left as a free parameter [33], and is chosen to best match the experimental data. For $U = 3.85\ k_BT$, we choose $\tilde{\phi}$ within the range for which MCT predicts nonergodicity. For $U = 2.89\ k_BT$ and $U = 3.06\ k_BT$, we choose values of $\tilde{\phi}$ near to the nonergodic transition, but still within the ergodic regime. We evaluate $f_c(q)$ by choosing a time for which the calculated value matches that of the experiment at $qa = 3.5$. For all $U$, our MCT predictions are in excellent agreement with the data, as shown by the solid lines in Fig. 4; moreover, this good agreement persists if we evaluate $f_c(q)$ at longer times, where $f(q,t)$ has decayed further. This agreement is remarkable considering the wide separation between network and nearest-neighbor length scales, and the highly heterogeneous structures. It highlights the critical role of glassy dynamics in kinetic arrest of the spinodal decomposition, leading to formation of a solid network [13].

This MCT calculation captures the increase of the non-ergodic parameter at low wave vectors, indicative of the presence of the peak in $S(q)$ at $qa < 0.2$ [31]. This increase reflects the constraints on motion at these length scales due to the existence of the interconnected space-spanning network. If the low-$q$ peak is not included, the predicted $f_c(q)$ is significantly lower, as shown by the dashed lines in Fig. 4. However, we find that the transition is still predicted to occur at nearly the same local volume fraction when the low-$q$ peak is excluded. This implies that the large-length-scale structures do not drive the arrest as predicted for a cluster glass transition [12, 22]; instead, gelation is driven by local arrest of the dynamics in a preformed, connected network. To further test for a possible arrest due to the formation of a glass of clusters each comprising many particles, we recalculate the MCT using as input only the low-$q$ network peak in $S(q)$. This leads to gelation at unphysically high particle densities, thus excluding this scenario [34].

For the systems studied here, gelation is driven by spinodal decomposition into an interconnected colloid-rich network, followed by dynamic arrest due to a local glass transition. By contrast, at higher $U$, and lower $\phi$, there is a state composed of disconnected glassy clusters, which is a precursor to gelation [6]. Such a fluid-cluster state may result from the effects of gravitational sedimentation or charge on the particles [35-37]. Alternatively, these glassy clusters could result from spinodal or binodal decomposition into a phase of disconnected droplets which are then locally arrested by the glass transition, but which do not span space, and hence do not form a gel [12]. The full range of scenarios for the gelation transition depends on $U$, $\xi$ and $\phi$, as well as possible influences of gravity and charge. The results presented here account for one scenario for gelation and serve as a benchmark to compare to other possible mechanisms.

We thank J. Bergenholtz, K. Kroy, M.E. Cates, F. Sciortino and B. Halperin for helpful discussions. This work was supported by NASA (NAG3-2284), NSF (DMR-0243715 and CHE-0134969 [DR and KM]), and the Harvard MRSEC (DMR-0213805).

HW and VT acknowledge support from the Swiss National Science Foundation.

**References**

1. S. M. Ilett, A. Orrock, W. C. K. Poon, and P. N. Pusey, *Phys. Rev. E* **51**, 1344 (1995).

2. W. C. K. Poon, F. Renth, R. M. L. Evans, D. J. Fairhurst, M. E. Cates, and P. N. Pusey, *Phys. Rev. Lett.* **83**, 1239 (1999).

3. W. van Megen and P. N. Pusey, *Phys. Rev. A* **43**, 5429 (1991).

4. K. N. Pham, A. M. Puertas, J. Bergenholtz, S. U. Egelhaaf, A. Moussaïd, P. N. Pusey, A. B. Schofield, M. E. Cates, M. Fuchs, and W. C. K. Poon, *Science* **296**, 104 (2002).

5. E. Bartsch, *Curr. Op. Colloid Interface Sci.* **3**, 577 (1998).

6. P. N. Segre, V. Prasad, A. B. Schofield, and D. A. Weitz, *Phys. Rev. Lett.* **86**, 6042 (2001).

7. W. Gotze, *in* "Liquids, Freezing, and the Glass Transition" (J. P. Hansen, D. Levesque, and J. Zinn-Justin, eds.). North-Holland, Amsterdam, 1991.

8. W. van Megen and S. M. Underwood, *Phys. Rev. E* **49**, 4206 (1994).

9. L. Fabbian, W. Gotze, F. Sciortino, P. Tartaglia, and F. Thiery, *Phys. Rev. E* **59**, R1347 (1999).

10. M. Carpineti and M. Giglio, *Phys. Rev. Lett.* **68**, 3327 (1992).

11. J. Bibette, T. G. Mason, H. Gang, and D. A. Weitz, *Phys. Rev. Lett.* **69**, 981 (1992).

12. M. E. Cates, M. Fuchs, K. Kroy, W. C. K. Poon, and A. M. Puertas, *J. Phys. Cond. Mat.* **16**, S4861 (2004).

13. E. Zaccarelli, F. Sciortino, S. V. Buldyrev, and P. Tartaglia, *in* "Proceedings of Unifying Concepts in Granular Materials and Glasses, Capri 2003" (A. Coniglio, A. Fierro, H. Hermann, and M. Nicodemi, eds.). Elsevier, Amsterdam, 2004.

14. S. Ramakrishnan, M. Fuchs, K. S. Schweitzer, and C. F. Zukoski, *J. Chem. Phys.* **116**, 2201 (2002).

15. N. A. M. Verhaegh, D. Asnaghi, H. N. W. Lekkerkerker, M. Giglio, and L. Cipelletti, *Physica A* **242**, 104 (1997).

16. K. G. Soga, J. R. Melrose, and R. C. Ball, *J. Chem. Phys.* **108**, 6026 (1998).

17. V. Prasad, V. Trappe, A. D. Dinsmore, P. N. Segre, L. Cipelletti, and A. D. Weitz, *Faraday Discuss.* **123**, 1 (2003).

18. I. Saika-Voivod, E. Zaccarelli, F. Sciortino, S. V. Buldyrev, and P. Tartaglia, *Phys. Rev. E* **70**, 041401 (2004).

19. E. Zaccarelli, S. V. Buldyrev, E. La Nave, A. J. Moreno, I. Saika-Voivod, F. Sciortino, and P. Tartaglia, *Phys. Rev. Lett.* **94**, 218301 (2005).

20. S. C. Glotzer, M. F. Gyure, F. Sciortino, A. Coniglio, and H. E. Stanley, *Phys. Rev. E* **49**, 247 (1994).

21. G. Foffi, C. de Michele, F. Sciortino, and P. Tartaglia, *Phys. Rev. Lett.* **94**, 078301 (2005).

22. K. Kroy, M. E. Cates, and W. C. K. Poon, *Phys. Rev. Lett.* **92**, 148302 (2004).

23. A. Yethiraj and A. van Blaaderen, *Nature* **421**, 513 (2003).

24. S. Asakura and F. Oosawa, *J. Chem. Phys.* **22**, 1255 (1954).

25. A. Vrij, *Pure Appl. Chem.* **48**, 471 (1976).

26. J. K. Percus and G. J. Yevick, *Phys. Rev.* **110**, 1 (1958).

27. A. E. Bailey and D. S. Cannell, *Phys. Rev. E* **50**, 4853 (1994).

28. E. O. Potma, D. J. Jones, J.-X. Cheng, X. S. Xie, and J. Ye, *Opt. Lett.* **27**, 1168 (2002).

29. J.-X. Cheng and X. S. Xie, *J. Phys. Chem. B* **108**, 827 (2004).

30. P. N. Pusey and W. van Megen, *Physica A* **157**, 705 (1989).

31. A. M. Puertas, M. Fuchs, and M. E. Cates, *J. Phys. Chem. B* **109**, 6666 (2005).

32. See online supplementary material, www.prl.aps.org

33. W. Kob, M. Nauroth, and F. Sciortino, *J. Non-Cryst. Solids* **307-310**, 181 (2002).

34. K. Miyazaki, *unpublished*.

35. J. Groenewold and W. K. Kegel, *J. Phys. Chem. B* **105**, 11702 (2001).

36. H. Sedgwick, K. Kroy, A. Salonen, M. B. Roberts, S. U. Egelhaaf, and W. C. K. Poon, *Eur. Phys. J. E* **16**, 77 (2005).

37. F. Sciortino, S. Mossa, E. Zaccarelli, and P. Tartaglia, *Phys. Rev. Lett.* **93**, 055701 (2004).

**Figure Captions:**

**Figure 1**. Wave vector dependence of the static structure factor for $U/k_BT = 2.46$ [--], 2.62 [···], 2.89 [○], 3.06 [△], and 3.85 [◇]. The solid line corresponds to the hard-sphere $S(q)$ expected at $\phi = 0.25$. Inset: $U$-dependence of $\Delta S_m$ defined as the difference between the value of $S(q)$ at the nearest neighbor peak and that at the minimum at $qa \sim 1$. The dotted line denotes the point at which the particle-particle position becomes highly correlated. A sharp peak appears concurrently at low $q$.

**Figure 2**. Images obtained by CARS microscopy: $U/k_BT = 2.46$ [top left]; $U/k_BT = 2.89$ [top right]; $U/k_BT = 3.06$ [bottom left]; $U/k_BT = 3.85$ [bottom right]. The dotted line between the two upper images denotes the boundary beyond which long lasting, space-filling structures are observed. The scale-bars correspond to 10 μm and the circles to $\pi/q_c$, where $q_c$ is obtained from the small angle static light scattering data. Insert: Schematic phase diagram for colloid-polymer mixtures as polymer concentration and colloid volume fraction are varied, where GL is the glass transition line. Points denote the approximate location of our samples.

**Figure 3**. (a) Dynamic structure factor at the $q$-value of the nearest neighbor peak for $U/k_BT = 2.46$ [--], 2.62 [···], 2.89 [○], 3.06 [△], and 3.85 [◇]. To account for the increased viscosity due to the added polymer, the time axis is normalized by the ratio of the viscosity of the polymer solution to that of the solvent, $\eta/\eta_0$. (b) Dynamic structure factor for $U/k_BT = 2.89$ obtained at $qa = 1.29$ [□], 1.9 [|], 2.49 [△], 3.03 [○], 3.52 [▽],

4.07 [—], and 4.31 [×]. The vertical line denotes the $f(q,t)$ values used to approximate the nonergodicity factor $f_c(q)$.

**Figure 4**. Wave vector dependence of the nonergodicity factor. The symbols are the measured $f_c(q)$. From bottom to top the data are for $U/k_BT = 2.89$, 3.06 and 3.85. The solid lines are obtained from the MCT calculation using $S(q)$ as input; experimental $S(q)$ at low and high $q$ were smoothed and joined with a spline fit, while at the highest $q$ we use an asymptotic expression for the numerical solution of the PY equation for the Asakura-Oosawa potential, adjusted to match the data. We use values $\tilde{\phi} = 0.37$, 0.52 and 0.50 respectively; the dashed lines are obtained by neglecting the low angle peak.

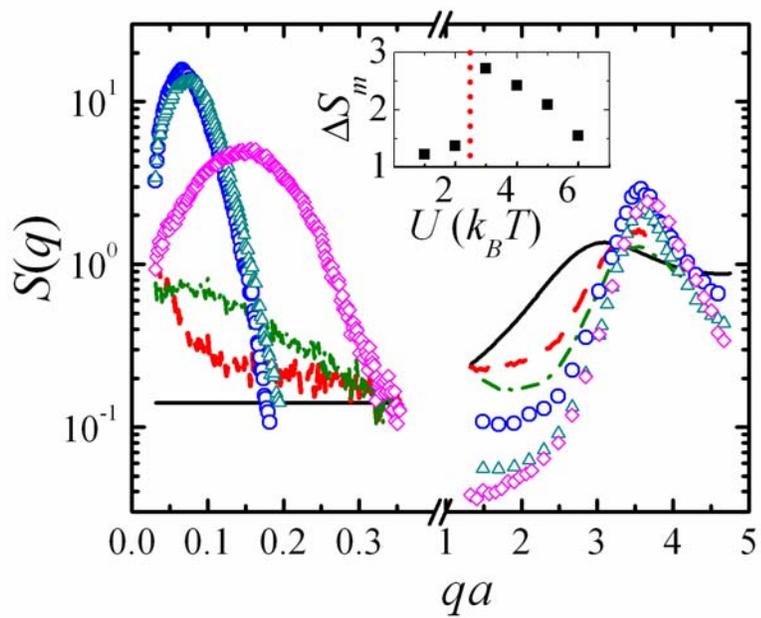

Fig. 1

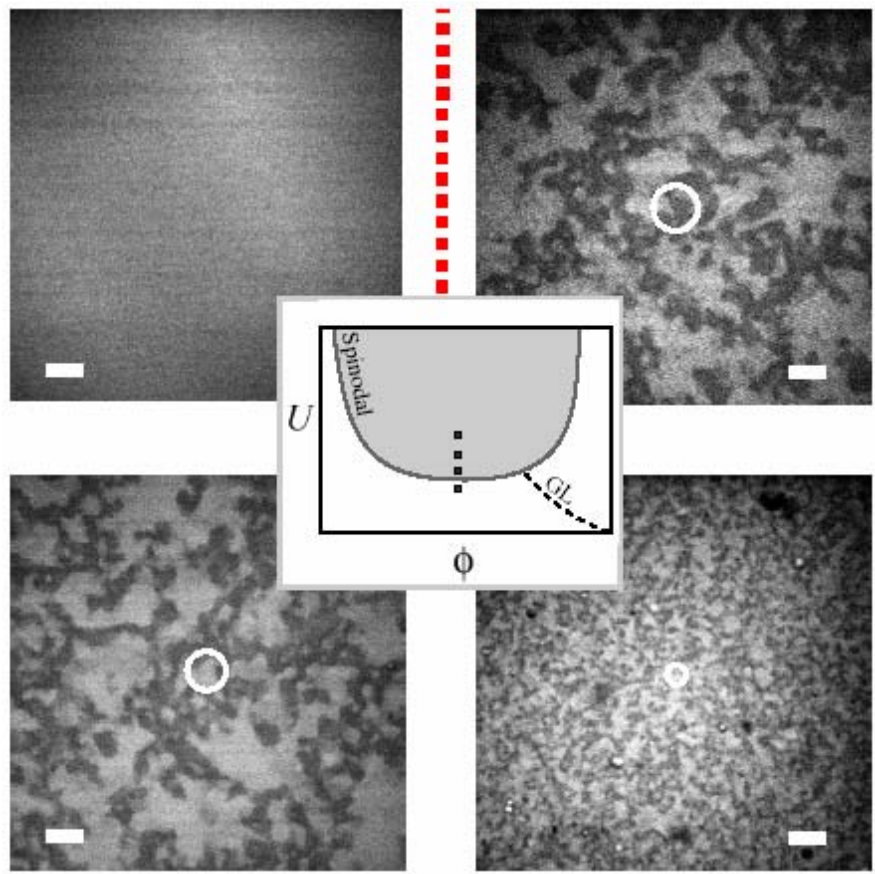

Fig. 2

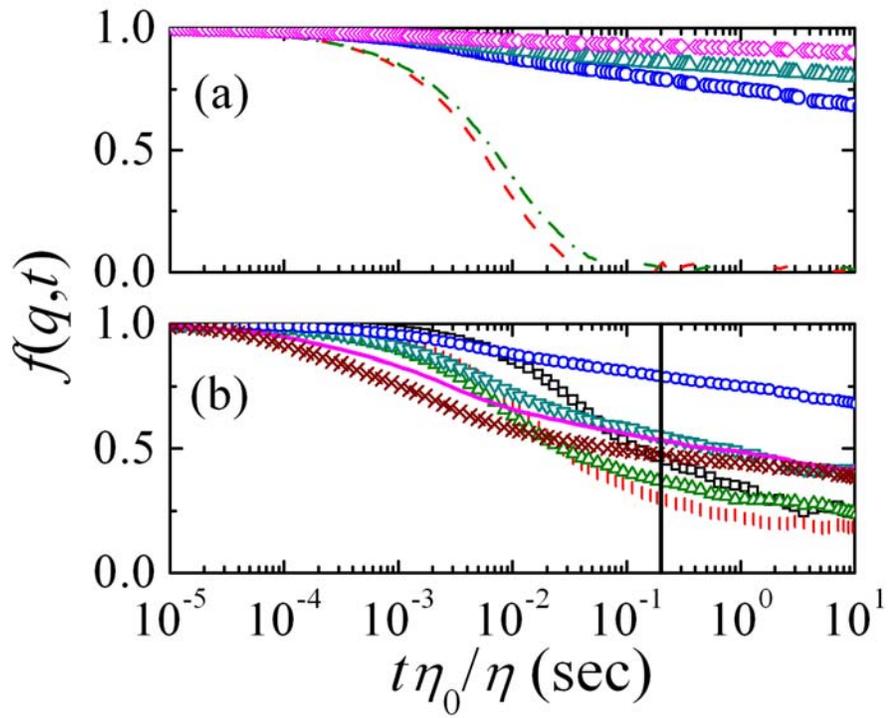

Fig. 3

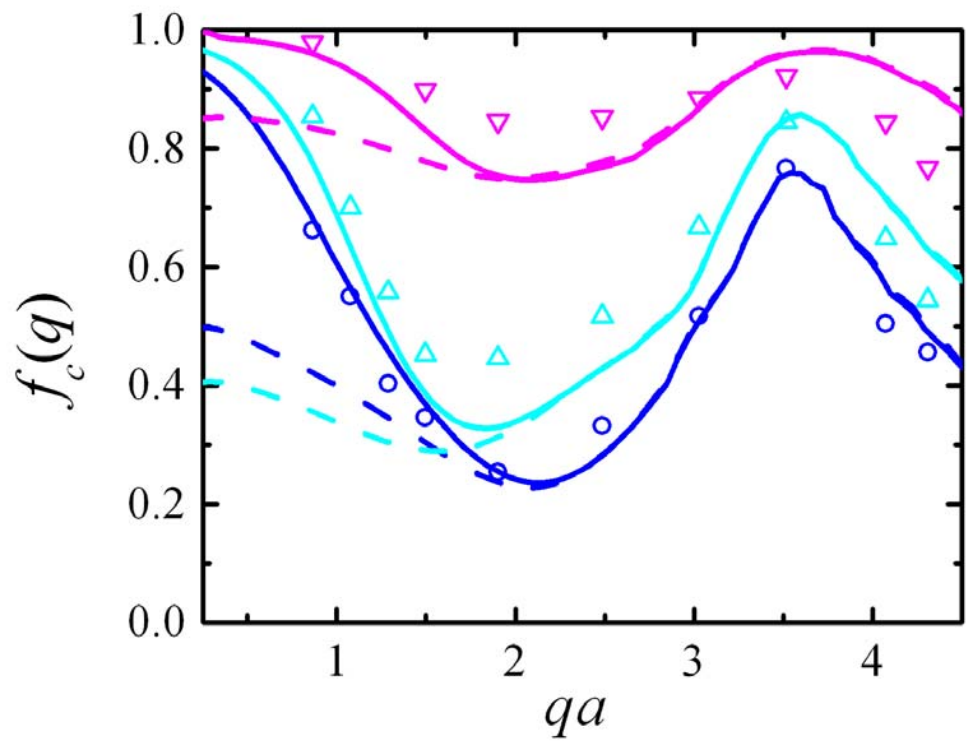